# $MgB_2$ conductors for dc and ac applications

B.A.Glowacki [a,b] and M.Majoros [b]

[a] Department of Materials Science and Metallurgy, University of Cambridge, Pembroke Street, Cambridge CB2 3QZ, UK,

[b] Interdisciplinary Research Centre in Superconductivity, University of Cambridge, Madingley Road, Cambridge CB3 OHE, UK

Abstract

The paper presents discussion on up to date results on $MgB_2$ conductors from the point of view of their future dc and ac applications. Basic physical parameters of $MgB_2$ compound relevant to conductors are introduced. Different conductor preparation methods and conductor architectures are presented and attainable critical current densities discussed. Some numerical results on critical currents and ac losses of future multifilamentary $MgB_2$ conductors with magnetic cladding of their filaments are given. Recently observed anomalous decrease of ac susceptibility at 50 K in copper claded Powder-in-tube, PIT, $MgB_2$ wires is presented.

*Key words*: superconducting conductors, $MgB_2$ superconductors, critical current, processing, composites, critical temperature anomaly.

*B.A.Glowacki , Department of Materials Science and Metallurgy, University of Cambridge, Pembroke Street, Cambridge CB2 3QZ, UK, fax 00441223334567  e-mail bag10@cam.ac.uk*

## 1. Introduction

Of the 96 binary boride systems known, 60 have compounds of boron with metals of the Ia-VIIIa subgroups crystallising in more than 30 structural types. The stability of borides formed with metals decreases with increasing atomic number, and borides of copper do not exist. Cu-B, Ag-B and Au-B systems are simple eutectic alloys [1].

The recent discovery of superconductivity in the well known $MgB_2$ material [2] makes all of us aware of the possible existence of new superconducting materials which may have critical parameters in some cases even better than existing superconductors. As it was described in New Scientist '*In just a few months, a new superconductor has upstaged its rivals*'[3]. This important discovery stimulated two types of research: first towards a combinatorial search for new superconducting materials, and secondly optimisation of thermo-mechanical procedures for the manufacture of powder-in-tube, PIT, and coated conductors using doped and substituted Mg-B compounds. In this paper we will focus on the second approach where research will be conducted on development and manufacture of conductors for dc and ac energy device applications such as MRI, NMR, SMES magnets and transformers.

Commercial exploitation of the recently discovered $MgB_2$ superconductors will be severely limited unless mechanically robust, high critical current density, composite conductors can be fabricated with uniform properties over long lengths. Our experience with LTS such as the A15 conductors $Nb_3Sn$, $Nb_3(Al,Ge)$ and also the whole range of HTS conductors, suggests that to be able to exploit the full superconducting potential of the metal-borides, one has to address fundamental questions concerning the practical capability of the newly discovered material to conduct intergranular supercurrents in high magnetic fields. The published critical field data for the first $MgB_2$ flux grown single crystal show that there is a noticeable anisotropy, $\gamma$, manifested in $H_{c2}(0)$ and $H_{c1}(0)$ and $\xi(0)$ [4,5]. Research conducted on PIT conductors suggests that this observed low degree of anisotropy should not have serious consequences for the transport current capabilities of high current $MgB_2$ conductors.

## 2. Conductor Development routes and requirements

There are two different methods of conductor manufacture: an *in situ* technique in which an Mg+2B mixture can be used as a central core of the PIT conductor or as an external coating and reacted *in situ* to form $MgB_2$, and an *ex situ* technique in which fully reacted $MgB_2$ powder, which may be doped or chemically modified, can be used to fill the metal tube. In the case of *ex situ* coated conductors sintering may not be even required if the additional densification process of the multilayer sandwich structure will results in a fully interconnect final $MgB_2$ material.

*2.1 Ex situ process: Mechanically induced intergrain connectivity, deformation and densification in $MgB_2$ conductors*

It should be stressed that the compaction of the $MgB_2$ powder followed by swaging, drawing and rolling in metallic tubes does not cause any problems and the deformation is similar to that of graphite [6]. For the manufacture of $MgB_2$ conductors by the PIT technique we have adopted the advanced stainless-steel cladding technique that was introduced by our laboratory to make the first $YBa_2Cu_3O_{7-x}$ oxide superconducting conductors [7]. A similar powder-in-tube approach has recently been used to fabricate metal-clad $MgB_2$ wires using various metals, such as Cu, Ag, Ni, Cu-Ni, or Nb [8-13].

From a practical point of view, it is much better to manufacture round wires than tapes, although it appears that the rolling procedure densifies further the internal $MgB_2$ core, improving intergrain connectivity and increasing the critical current capabilities of the conductor. Increasing values of $J_c$ for the subsequently increasing degree of deformation of the superconducting core, as presented in Fig.1, underline the necessity to define the role of the mechanical deformation and slip planes and shearing forces on the intergranular connectivity of this unusual compound which may have some similarities with intercalated graphite. It appears that additional post deformation annealing helps to further improve the critical current density of the final conductor. Critical current measurements of mechanically strained $MgB_2$ tapes prove that there is an initial increase in $I_c$, similar to $Nb_3Sn$ and $Nb_3Al$ under bending stress followed by degradation due to transversal cracking [14].

Further understanding of the mechanism responsible for the improvement of the $J_c$ value and 'boundaryless' behaviour of the deformed $MgB_2$ crystallites is required. In this layered structure, boron atoms form honeycomb layers alternately with hexagonal layers of Mg atoms. Some analogies with intercalated graphite could be used to understand such improved transport behaviour.

*2.2. In situ process: Reactive diffusion formation of the $MgB_2$ phase*

The alternative approach, an *in situ* PIT wire made from fine stoichiometric powders such as B and Mg, is expected to be an equally good solution, however, deformation and solid-liquid diffusion processes are going to be complex and difficult, see Fig.2. This is firstly because wires manufactured by the *in situ* technique would experience ~25 % decrease in density, due to the phase transformation from $Mg+2(-B)=>MgB_2$, and secondly because of the stability of the higher borides at lower temperatures. The fact that B has hardness $H_v = 49000$ second only to diamond has significant consequences on uniformities of the multifilamentary wires in the case of *in situ* process.

During the *in situ* process the deformation of the B particles in the relatively soft Mg environment is not the best option for the multifilamentary wires where the initial particle size will in many cases define the diameter of the final filament. It may appear that the *in situ* technique can only be used with very fine B initial powder diameter, however fragmentation of some B particles has been observed. The finer the boron particles the better the interconnectivity between the grains in the final *in situ* conductor and the better the current percolation [15].

In the case of the *in situ* process in copper tubes lower reaction temperatures will be preferred in order to prevent the extensive rapid formation of Cu-Mg alloy which may affect chemical uniformity of the superconducting core and conductivity of the stabilising Cu layer. To prevent diffusion of Mg into copper one my use diffusion barriers such as Ni, Ta, Nb, or Fe. There are confirmed $T_c$ anomalies observed in Cu and Ag PIT *in situ* wires, Fig.3 which will be further researched [5, 16].

*2.3 Conductors for AC applications with improved critical current*

One can explore the possibilities of recently patented superconducting-magneting heterostructures [17, 18] with the aim of minimising ac losses for existing ac applications such as specific transformers or novel electrotechnology devices. Computer modelling was used to provide the base for initial assessment of the optimum design for given power applications.
Generally the ac losses in multifilamentary wires can be reduced by twisting their filaments, Fig.4. The shorter the twist pitch length the larger the ac loss reduction. The minimum practical twist pitch is approximately five times the diameter of the composite.
While these twist pitches are fully effective in uniform external magnetic fields, they are only partially effective in non uniform fields and less effective with respect to self-field of the composite. In self-field conditions the twist does not change the self-field flux linked between the inner and outer filaments substantially and the current first fills the outer layers of the superconducting composite, similar to a solid superconductor. To decouple the filaments in self-field conditions a magnetic screening method was proposed [17, 18]. This consists of surrounding each filament by a thin ferromagnetic layer.
Compared with other superconductors, $MgB_2$ has an advantage that it does not react with iron and moreover iron is the unique metal which can be used as a protective layer between $MgB_2$ and a stabilising outer copper layer [19]. So it automatically provides the magnetic shield of the filaments as well. It has been found that a multiple layer shield is more effective than a thicker single layer [20]. In Fig.5 we present a numerical estimate of hysteresis loss in a perpendicular applied magnetic field in a monocore $MgB_2$ wire. Its diameter was 1mm and it was shielded by 4 iron layers, each 50µm thick with 50µm normal metal layer separation. Laboratory iron with non linear *B(H)* characteristic and $\mu_{r(max)}$=9000 was used in calculations. Hysteresis loss in the iron shields is negligibly small in comparison with the loss in the $MgB_2$ wire. The losses in the shielded $MgB_2$ wire are negligibly small as compared with an unshielded one, up to a applied magnetic field of 0.4T, Fig.5. Magnetic fields of this magnitude appear in some applications as e.g. superconducting transformers. We have established by numerical modelling that magnetic shielding of the filaments reduces ac losses in self field conditions as well due to decoupling of the filaments and at the same time it increases the critical current of the composite. An example of the critical current increase in the case of shielded 19 filament composite is shown in Fig.6. The critical current increased from about 440 A up to about 630 A in self magnetic field. This is due to rather strong magnetic field dependence of the critical current density in $MgB_2$ material.

*2.4 Magnetic and cryogenic stability*

Due to the possible relatively large filament size in the new $MgB_2$ conductors there are two important aspect of quench protection and stability: magnetic and cryogenic. Firstly, to avoid flux jumps in the large filament $MgB_2$ conductors use of a magnetic shielding material such as iron is advantageous to reduce large magnetic flux gradient in the superconducting filaments (see Fig.6); secondly, to provide adequate cryogenic stabilisation an external highly conductive (thermally and electrically) metal is essential.

Complex modelling and measurements was conducted (using existing commercial software and measurement equipment) on the thermal and magnetic stability of magnetic and non-magnetic clad PIT conductors. It is also expected that during magnet winding procedure some defects can be induced such as microcracks. The results of the current transfer around the crack as well as the temperature distribution are shown in Fig 7a & 7b respectively. They show a sufficient cryogenic stability of the wire. The current passes through the copper and iron cladding without any noticeable overheating. The transfer length of the current through the normal metal cladding is approximately 1.6 mm from each side of the defect. The middle of the wire has the temperature increase about 7 mK. Maximum temperature increase is on the surface of the defect (about 12 mK) and the temperature increase on the surface of the copper cladding is about 0.5 mK a value well within the range of convectional heat flow of liquid helium.

## 3. Conclusions

The aim of the future research is to develop, through *ex-situ* and *in situ* thermo-mechanical processing, doped/substituted $MgB_2$ conductors and coils with maximised superconducting parameters for particular dc and ac applications such as NMR, MRI, cryogen free persistent mode magnets, specific transformers and SMES.

There is potential, and also a strong hope, that new reliable superconducting joints and superconducting switches will be made not only based on the new $MgB_2$ conductors but also between $MgB_2$ and low temperature superconducting conductors such as NbTi, $Nb_3Sn$ and $Nb_3Al$, opening a new range of the electromagnetic device applications.

**Acknowledgement**

M.Majoros acknowledges the AFRL/PRPS Wright-Patterson Air Force Base, Ohio, USA for the financial support.

## References

[1] T. Serebryakova, Journal of the Less-Common Metals, 67(1979)499.
[2] J. Nagamatsu, N. Nakagawa, T. Muranaka, Y. Zenitani, J. Akimitsu, Nature 410(2001)63.
[3] M. Cooper, 'Current champion' Frontiers Emerging Technologies, New Scientist, 2 June 2001 No.2293(2001)21.
[4] M. Xu, H. Kitazawa, Y. Takano, J.Ye, K. Nishida, et al. cond.-mat./0105271.
[5] B.A. Glowacki and M. Majoros in Studies of superconductors (Advances in Research and Applications)'*$MgB_2$ superconductors*', ed. A.Narlikar, Nova Science Publishers, Inc., Huntington, New York, 2001, v38. pp. in press.
[6] B.A. Glowacki and Z.M. Kosek, Cryogenics 27 (1987)551.
[7] B.A. Glowacki and J.E. Evetts, Mat. Res. Soc. Symp. Proc., 99(1988)419.
[8] B.A. Glowacki, M. Majoros, M. Vickers, J.E. Evetts, Y. Shi, I. McDougall, Supercond. Sci. Technol. 14 (2001) 193.
[9] M.D. Sumption, X. Peng, E. Lee, M. Tomsic and E.W. Collings, cond- mat/ 010244.
[10] G. Grasso, A. Malagoli, C. Ferdeghini, et al., Appl. Phys. Lett., 79 (2001) 230.
[11] W. Goldacker, S.I. Schlachter, S. Zimmer, H. Reiner, cond-mat/0106226.
[12] X.L. Wang, S. Soltanian, et al., con-mat 0106148.
[13] H-L. Suo, C.Beneduce, M.Dhallé, N.Musolino, X-D. Su, E. Walker, P.Toulemonde and R.Flükiger, Applied Physics Letters in press.
[14] H. Kitaguchi, H. Kumakura and K. Togano cond-mat/0106388.
[15] N.A. Rutter and B.A. Glowacki, IEEE Trans. Applied Superconductivity, 11(2001)2730.
[16] B.A.Glowacki, M.Majoros, M.E.Vickers and B.Zeimetz, EUCAS, 26-30 August, 2001 Copenhagen, Denmark.
[17] B.A. Glowacki and M. Majoros, Supercon. Sci. Techn., 13(2000)971.
[18] M. Majoros, B.A. Glowacki and A.M. Campbell, Physica C, 334(2000)129.
[19] S. Jin, H. Navoori, C. Bower and R.B.van Dover, Nature, 411(2001)563.
[20] M. Majoros, B.A. Glowacki et al. IEEE Trans. Applied Supercond., 11(2001)2780.

## Figure captions

Fig.1 Schematic outline of the *ex situ* $MgB_2$ conductor manufacture.

Fig.2 Schematic outline of the *in situ* $MgB_2$ conductor manufacture.

Fig.3 The 50K AC susceptibility anomalies of *in situ* Cu-Mg-B wire sintered at 700°C for 1hour under protective argon atmosphere. The inset shows the full curve in the temperature range from 5K to 100K [5]; transport $J_c > 10^5 Acm^{-2}$.

Fig.4 A *kA-class* multifilamentary $MgB_2$ conductor for ac and dc applications; 6 x Cu/Mg-2B *in situ* filaments twisted around a central stainless steel core, the inset represents an individual single core conductor, diameter 0.5mm. [5].

Fig.5 AC losses of a single core conductor in an external magnetic field; ○- unscreened conductor ● -magnetically screened conductor with 4 concentric separated iron layers.

Fig.6 Spatial distribution of the critical current density in a 19 filament $MgB_2$ wire cross section in self field, for different values of relative magnetic permeability, $\mu_r$, of the concentric multi-screens: a) $\mu_r=1$, $I_c=442A$, b) Non-linear Fe $\mu_{rmax}=9000$, $I_c=628A$ [5].

Fig.7 The upper half of the longitudinal cross-section of the Cu-Fe-$MgB_2$ wire (diameter 2mm, $J_c=2.65109 Am^{-2}$): a) current flow along the wire around the defect (z-component), b) temperature distribution in the longitudinal wire cross-section around the defect.

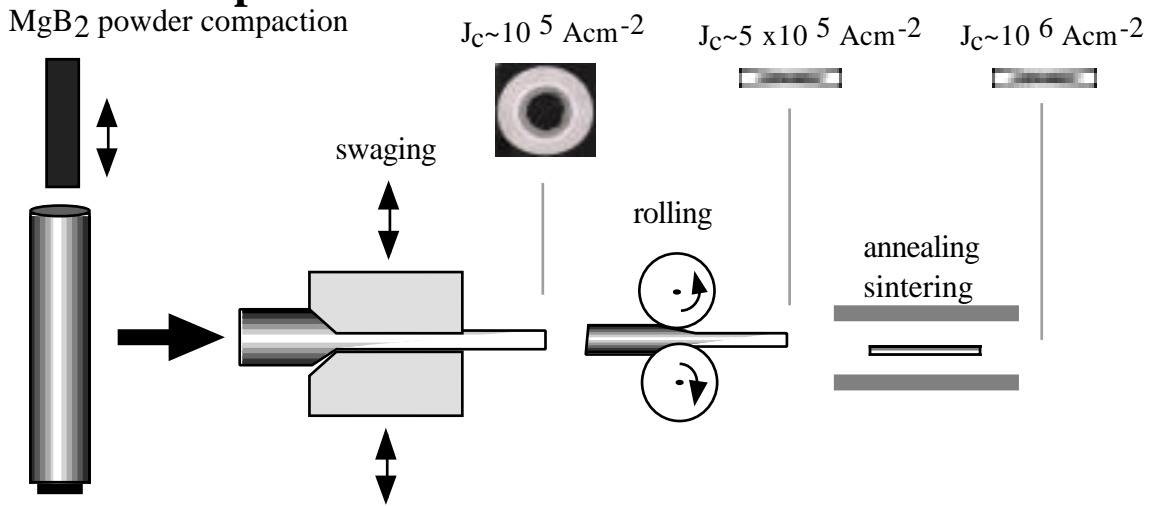

Fig.1

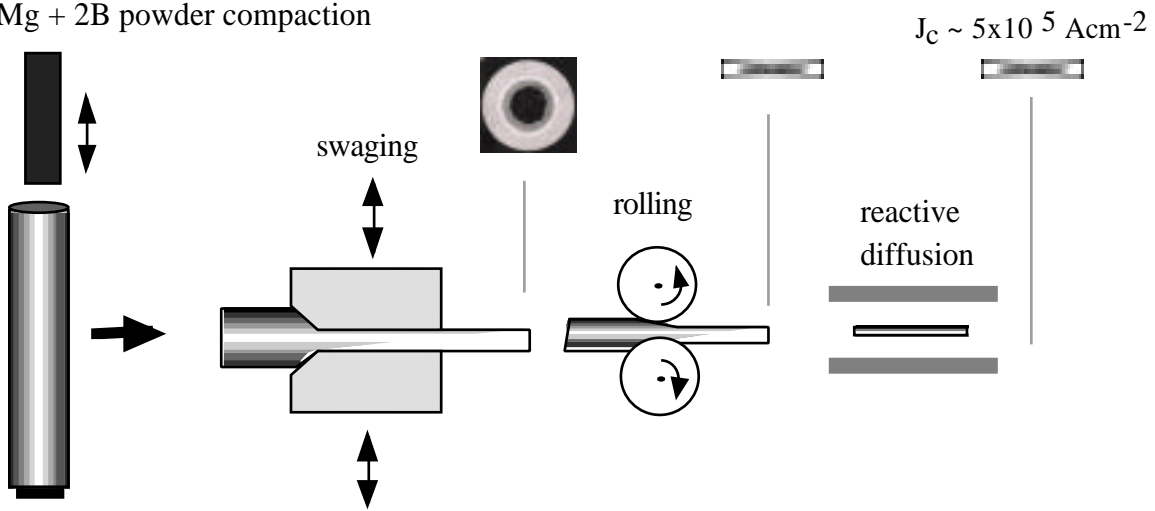

Fig.2

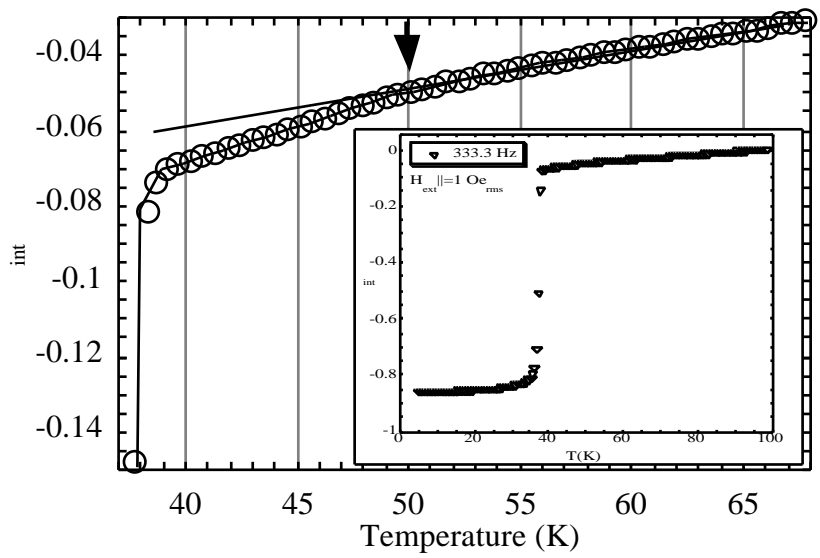

Fig.3

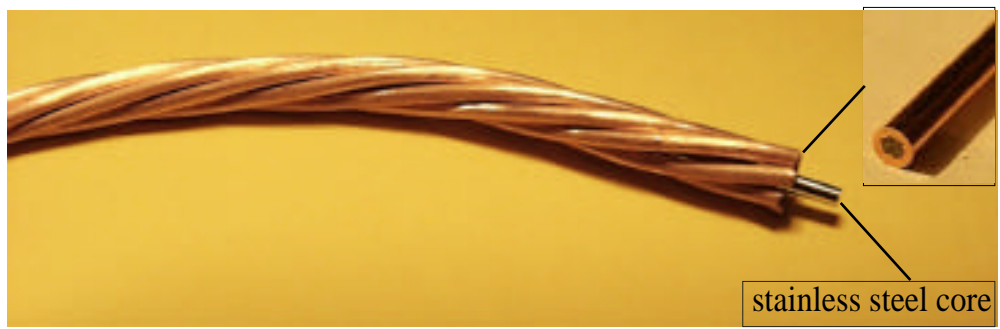

Fig.4

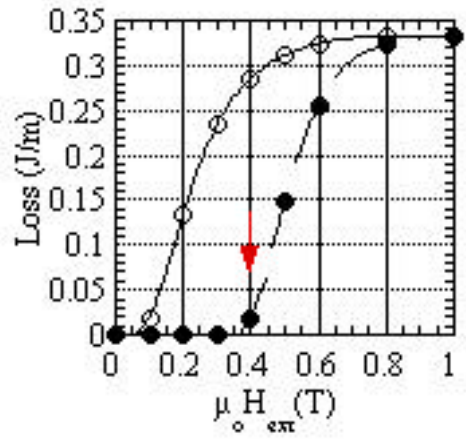

Fig.5

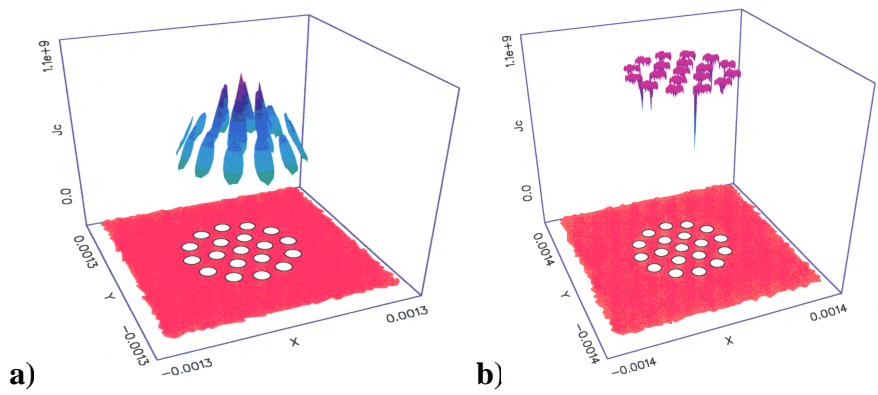

Fig.6

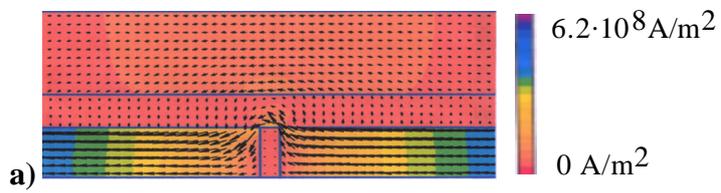

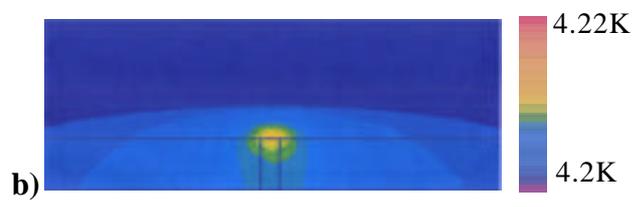

Fig.7